\documentclass[prl,twocolumn,lengthcheck]{revtex4}

\usepackage{amsmath}
\usepackage{amsthm}
\usepackage{amssymb}
\usepackage{graphicx}

\newcommand{\uinvnorm}{|\kern-2pt|\kern-2pt|}

\newcommand{\dist}{\operatorname{dist}}
\newcommand{\spectrum}{\operatorname{spec}}

\theoremstyle{plain}
\newtheorem{theorem}{Theorem}

\newtheorem{proposition}[theorem]{Proposition}

\theoremstyle{definition}
\newtheorem{definition}[theorem]{Definition}
\newtheorem{example}[theorem]{Example}

\theoremstyle{remark}
\newtheorem{remark}[theorem]{Remark}

\begin{document}
\bibliographystyle{apsrev}

\title{Approximate locality for quantum systems on graphs}

\author{Tobias J.\ Osborne}
\email[]{Tobias.Osborne@rhul.ac.uk} \affiliation{Department of
Mathematics, Royal Holloway University of London, Egham, Surrey TW20
0EX, United Kingdom}

\date{\today}

\begin{abstract}
In this Letter we make progress on a longstanding open problem of
Aaronson and Ambainis [Theory of Computing {\bf 1}, 47 (2005)]: we
show that if $A$ is the adjacency matrix of a sufficiently sparse
low-dimensional graph then the unitary operator $e^{it{A}}$ can be
approximated by a unitary operator $U(t)$ whose sparsity pattern is
exactly that of a low-dimensional graph which gets more dense as
$|t|$ increases. Secondly, we show that if $U$ is a sparse unitary
operator with a gap $\Delta$ in its spectrum, then there exists an
approximate logarithm $H$ of $U$ which is also sparse. The sparsity
pattern of $H$ gets more dense as $1/\Delta$ increases. These two
results can be interpreted as a way to convert between local
continuous-time and local discrete-time processes. As an example we
show that the discrete-time coined quantum walk can be realised as
an approximately local continuous-time quantum walk. Finally, we use
our construction to provide a definition for a fractional quantum
fourier transform.
\end{abstract}

\maketitle

%
%

In physics the word \emph{locality} admits many possible
interpretations. In quantum field theory and condensed matter
physics locality is understood as the \emph{clustering of
correlations} \cite{endnote18, weinberg:1996a, peskin:1995a,
fredenhagen:1985a, lieb:1972a, hastings:2004a, cramer:2006a}. In
quantum information theory the \emph{quantum circuit model}
\cite{nielsen:2000a, preskillnotes} reigns supreme as the final
arbiter of locality where it is natural to define the
\emph{nonlocality} of a physical process to be the minimal number of
fundamental two-qubit quantum gates required to simulate the process
up to some prespecified error \cite{endnote19}. The central role the
quantum circuit model plays in assessing the nonlocal ``cost" of a
physical process strongly motivates us to quantify the relationship
between the notions of locality accepted in other branches of
physics and the quantum gate cost of quantum information theory.

Thus, we appear to have at least two different interpretations of
the word locality for quantum systems: on one hand we have the
clustering of many-particle physics, and on the other we have the
gate cost of quantum information theory. From a physical perspective
it is in intuitively clear that there should be a strong
relationship between these two definitions. After all, dynamical
clustering implies a bound on the speed of information transmission.
Indeed, for many particle systems there are now results quantifying
this relationship: a low-dimensional system which exhibits dynamical
clustering can be simulated by a constant-depth quantum circuit
\cite{osborne:2005d}.

However, for the case of scalar and spinor quantum particles hopping
on finite graphs an explication of the connections between the
clustering-type interpretation of locality and the quantum-circuit
type interpretation has yet to be completed. An investigation of the
graph setting was initiated by Aaronson and Ambainis
\cite{aaronson:2005a}, who established the canonical analogues of
the clustering and quantum-circuit definitions of locality. The main
questions remaining are now to quantify the relationship between the
quantum-circuit locality (what Aaronson and Ambainis call
``$Z$-locality" and ``$C$-locality") and the clustering-type
interpretation (called ``$H$-locality") for these systems.

The objective of this Letter is to provide a (not necessarily
optimal) equivalence between the notions of locality introduced in
\cite{aaronson:2005a}, thus partially resolving one of their
longstanding conjectures: we establish that a graph-local
continuous-time process can be written (``discretised") as a product
of discrete-time processes (``quantum gates"). Conversely, we show
how to compute an approximately local logarithm for a unitary gate
which is local on some graph $G$. In other words, we show how to
construct a local continuous-time process $\mathcal{C}$ associated
with a local discrete-time quantum process $\mathcal{D}$ such that
$\mathcal{D}$ may be realised by sampling $\mathcal{C}$ at
appropriate intervals.

\begin{figure*}
\includegraphics{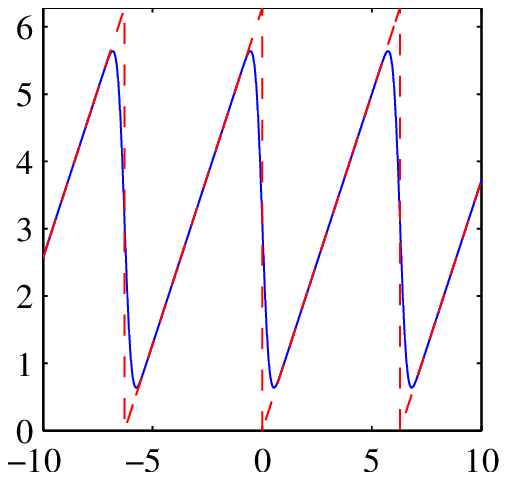}
\hspace{1cm}
\includegraphics{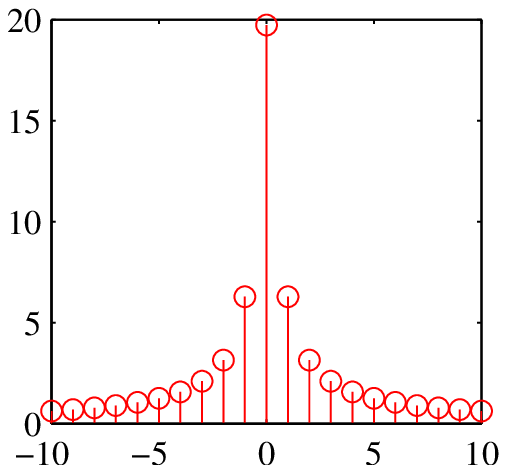}
\hspace{1cm}
\includegraphics{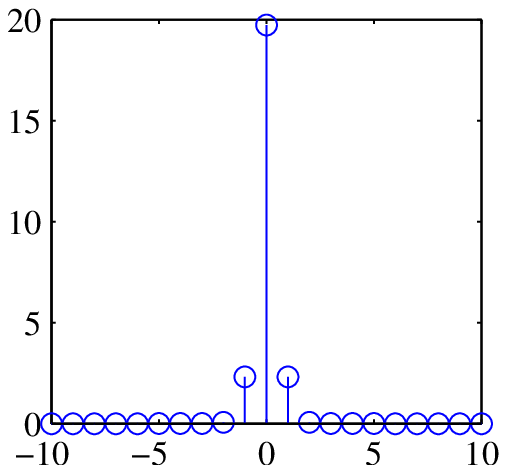}
\caption{Smoothing out the sawtooth to improve convergence of
fourier coefficients: in the first figure the original sawtooth wave
is shown in red, along with the smoothed version in blue. In the
second figure the absolute values of ($2\pi$ times) the fourier
coefficients of the original sawtooth are shown. Note the slow
convergence of this sequence. Finally, in the third figure the
fourier coefficients of the smoothed sawtooth are shown. Note the
improved convergence of the sequence.}\label{fig:smeartooth}
\end{figure*}

All the quantum systems we consider in this Letter are naturally
associated with a \emph{finite graph} $G = (V, E)$, where $V$ is a
set of $n$ \emph{vertices} and $E$ a set of \emph{edges}. We write
$v\sim w$ if $(v, w) \in E$. We summarise this connectivity
information using the \emph{adjacency matrix} $A$, which has matrix
elements given by $A_{v,w} = 1$ if $v\sim w$ and $A_{v,w} = 0$
otherwise. For two vertices $v, w \in V$ we let $\dist(v,w)$ denote
the graph-theoretical distance --- the length of the shortest path
connecting $v$ and $w$, with respect to the edge set $E$. Let $M \in
\mathcal{M}_{n}(\mathbb{C})$ be an $n\times n$ matrix. The
\emph{sparsity pattern} $A$ of $M$ is the $n\times n$
$\{0,1\}$-matrix given by: $A_{j,k} = 0$ if $M_{j,k} = 0$ and
$A_{j,k} = 1$ if $M_{j,k}\not=0$. It is sometimes convenient in the
sequel to arbitrarily assign directions (arrows) to the edges of
$G$. In this case we write $e^+$ (respectively, $e^-$) for the
vertex at the beginning (respectively, end) of $e$. Finally, we
denote by $D(G)$ the \emph{maximum degree} of $G$, which is the
maximum number of edges which are incident to any vertex in $G$.

There is a canonical way to associate a Hilbert space
$\mathcal{H}_V$ with a finite graph $G$ with vertex set $V$: we use
vertices to label a basis of quantum states, so that $\mathcal{H}_V
\equiv \langle |v\rangle \, | \, v\in V\rangle$
--- this is the Hilbert space of a scalar quantum particle
constrained to live on the vertices of $G$.

We now recall the definitions of locality introduced by Aaronson and
Ambainis \cite{aaronson:2005a} for a quantum particle on a graph.
Note that the definitions we present here are not as general as
those introduced in \cite{aaronson:2005a}: Aaronson and Ambainis
include the possibility of an extra internal degree of freedom.
While, for clarity, we ignore this extra internal degree of freedom
it is straightforward to extend our results to cover the more
general case.

\begin{definition}
A unitary matrix $U$ is said to be $Z$-local on $G$ if $U_{j,k} = 0$
whenever $j\not= k$ and $(j, k) \not\in E$.
\end{definition}

\begin{definition}
A unitary matrix $U$ is said to be $C$-local on $G$ if:
\begin{enumerate}
\item the basis states $|v\rangle$ can be partitioned into subsets
$\mathcal{P}_1, \mathcal{P}_2, \ldots, \mathcal{P}_q$ such that
$U_{j,k} = 0$ whenever $|j\rangle$ and $|k\rangle$ belong to
distinct subsets $P_l$; and %
\item for each $j$, all basis states in $\mathcal{P}_j$ are either
from the same vertex or from two adjacent vertices. %
\end{enumerate}
\end{definition}

\begin{definition}
A unitary matrix $U$ is said to be $H$-local on $G$ if $U = e^{iH}$
for some hermitian matrix $H$ with $\|H\|_\infty \le \pi$ such that
$H_{j,k} = 0$ whenever $j\not= k$ and $(j, k) \not\in E$.
\end{definition}

The first result we prove in this Letter shows how an $H$-local
unitary operator may be written as a product of $C$-local unitary
operators. This result is entirely standard and is a straightforward
corollary of the \emph{sparse hamiltonian lemma} of
\cite{aharonov:2003a}. We sketch a proof for completeness.

\begin{proposition}\label{prop:explocal}
Let $H$ be the adjacency matrix of a finite graph $G$. Then
$e^{itH}$ may be approximated by a product of $c|t|D(G)$ $C$-local
unitary operators, where $c$ is some constant. Because a product of
$C$-local unitary operators is $Z$-local on some graph related to
$G$ an $H$-local unitary operator is approximately $Z$-local on some
graph $G'$ related to $G$, which gets denser as $|t|$ increases.
\end{proposition}
\begin{proof}
The idea behind the proof is as follows. We first write $H=
\sum_{j=1}^{D(G)+1} h_j$, where $h_j = \sum_{e\in \mathcal{C}_j}
\alpha_e^{(j)}$, $\alpha_e^{(j)} = |e^+\rangle \langle e^-| +
\text{h.c.}$, and $[\alpha_e^{(j)}, \alpha_f^{(j)}]=0$ (this
decomposition follows from a colouring of the edges provided by
Vizing's theorem \cite{vizing:1964a}: we denote by $\mathcal{C}_j$
the set of edges with the same colour). Then we use the Lie-Trotter
formula to approximate $e^{itH}$ by powers of $U_\delta =
(e^{i\delta h_1}e^{i\delta h_2}\cdots e^{i\delta h_m})(e^{i\delta
h_m} \cdots e^{i\delta h_1})$:
\begin{equation}
\|U_{\delta}^{\lfloor \frac{|t|}{2\delta}\rfloor}- e^{itH}\|_\infty
\le O(m\Lambda\delta + m\Lambda^3|t|\delta^2),
\end{equation}
where $m=D(G)+1$ and $\Lambda = \max_j \|h_j\|_\infty \le 2$, where
the inequality for $\|h_j\|$ follows straightforwardly from, for
example, Ger\v{s}gorin's circle theorem \cite{horn:1990a}. Finally,
we observe that $e^{i\delta h_j}$ is a $C$-local unitary operator,
for each $j=1, 2, \ldots, m$.
\end{proof}

\begin{figure*}
\includegraphics{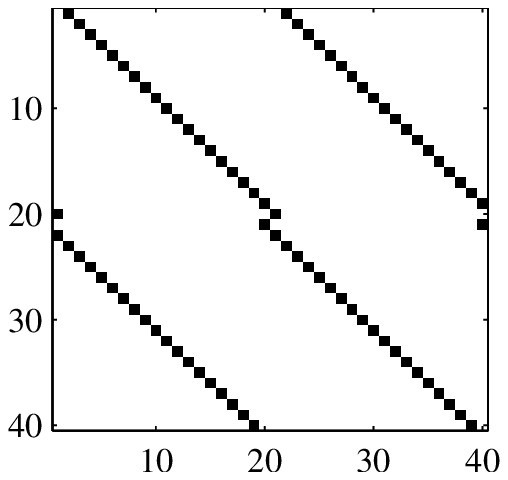}
\hspace{1cm}
\includegraphics{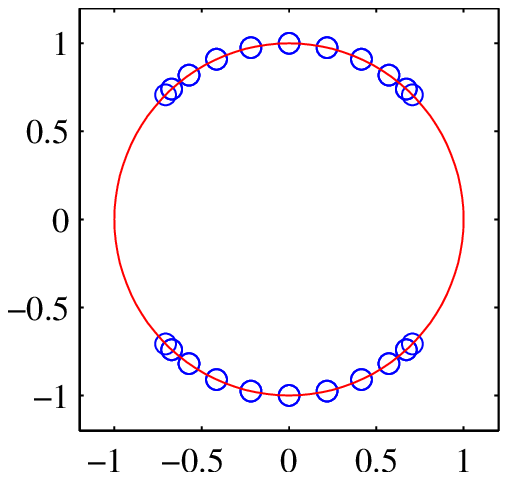}
\hspace{1cm}
\includegraphics{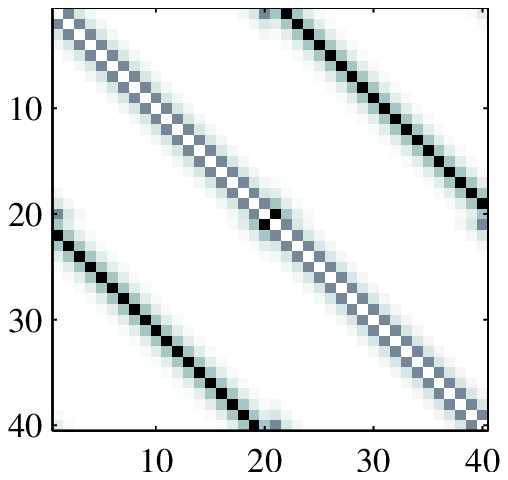}
\caption{The coined quantum walk $U$ on a ring with $20$ sites. The
first figure shows the sparsity pattern of $U$. The second figure
shows the spectrum of $U$ with a gap clearly evident. The third
figure shows the absolute values of the matrix elements of the
logarithm $H$ of $U$ constructed in the proof of
Proposition~\ref{prop:logu}. Note the rapid decay of the matrix
elements away from the original nonzero entries of
$U$.}\label{fig:hamiltonianme}
\end{figure*}

\begin{proposition}\label{prop:logu}
Let $U$ be a unitary matrix whose sparsity pattern $A$ is the
adjacency matrix of a digraph $G$. If the arguments $\theta$ of all
of the eigenvalues $e^{i\theta}$ of $U$ satisfy $\theta \in [0,
2\pi) \setminus (\alpha, \beta)$, with $\Delta = |\alpha - \beta|$,
then there exists a unitary matrix $V$ which is $H$-local on a graph
$G'$ given by the sparsity pattern of $A(G)^k$ where $k =
c/(\epsilon^2\Delta)$, for some constant $c$, such that
$\|U-V\|_\infty \le \epsilon$.
\end{proposition}
\begin{proof}

We begin by writing $U$ in its eigenbasis:
\begin{equation}
U = \sum_{j = 1}^{n} e^{i\phi_j} |j\rangle \langle j|,
\end{equation}
where $|j\rangle$ are the eigenvectors of $U$ and we choose $\phi_j
\in [0, 2\pi)$. By multiplying by an overall unimportant phase
$e^{i\zeta}\mathbb{I}$ we can set the zero of angle to arrange for a
gap in the spectrum $\spectrum(U)$ of $U$ to lie over the origin.
Such a gap always exists for finite dimensional unitary operators,
but not necessarily for infinite operators.

We want to find a hermitian matrix $H$ so that $U = e^{iH}$. We call
this the \emph{effective hamiltonian} for $U$. One such hamiltonian
is simply given by
\begin{equation}\label{eq:hlogu}
H = \sum_{j=1}^n \phi_j |j\rangle\langle j|.
\end{equation}
While this expression is perfectly well-defined, it is very hard to
see any kind of sparsity/local structure in $H$. To overcome this
we'll find an alternative expression for $H$ defined by
Eq.~(\ref{eq:hlogu}) as a power series in $U$. To do this we suppose
that
\begin{equation}\label{eq:hseries}
H = \sum_{k=-\infty}^{\infty} c_k U^k,
\end{equation}
and we solve for the coefficients $c_k$: we equate the coefficient
of $|j\rangle \langle j|$ on both sides to find
\begin{equation}
\phi_j = \sum_{k=-\infty}^{\infty} c_k e^{ik\phi_j}.
\end{equation}
Hence, if we can find $c_k$ such that
\begin{equation}\label{eq:ckseries}
\theta = \sum_{k=-\infty}^{\infty} c_k e^{ik\theta},
\end{equation}
for all $\theta \in [0, 2\pi)$ then we are done. (Recall that we've
arranged it so there are no eigenvalues of $U$ on the point $\theta
= 0$.) To solve for $c_k$ we integrate both sides of
Eq.~(\ref{eq:ckseries}) with respect to $\theta$ over the interval
$[0, 2\pi)$ against $\frac{1}{2\pi}e^{-il\theta}$, for
$l\in\mathbb{Z}$:
\begin{equation}\label{eq:ckseries2}
\frac{1}{2\pi}\int_{0}^{2\pi}\theta e^{-il\theta}\,d\theta =
\frac{1}{2\pi}\sum_{k=-\infty}^{\infty} c_k \int_{0}^{2\pi}
e^{i(k-l)\theta}\,d\theta.
\end{equation}
Thus we learn that the $c_k$ are nothing but the fourier
coefficients of the periodic sawtooth function $f(\theta+2\pi l) =
\theta$, $\theta \in [0, 2\pi)$, $l\in\mathbb{Z}$:
\begin{equation}\label{eq:cns}
c_k = \begin{cases}\pi, \quad k=0 \\ \frac{i}{k}, \quad
k\not=0.\end{cases}
\end{equation}

Now we know the formula for $c_k$ we substitute this into
Eq.~(\ref{eq:hseries}):
\begin{equation}
H = \sum_{k=-\infty}^{\infty}  \left(\pi\delta_{k,0} + \frac{i
(1-\delta_{k,0})}{k} \right)U^k,
\end{equation}
and truncate the series at some cutoff $k \le K$. If we assume the
sparsity pattern of $U$ describes a sufficiently sparse graph $G$
then $U^k$ will also describe a sparse graph for any constant $k$
\cite{endnote20}, and, as a consequence, the truncated series
representation for $H$ would also describe a sparse graph.

Unfortunately we cannot do this: the sawtooth wave has a jump
discontinuity and hence the fourier series is only conditionally
convergent. Thus it is impossible to truncate the series without a
serious error.

The way to proceed is to assume that we have some further
information, namely, that $U$ has a gap $\Delta$ in its spectrum.
The eigenvalues of $U$ lie on the unit circle in the complex plane
so what we mean here is that there is a continuous arc in the unit
circle which subtends an angle $\Delta$ where there are no
eigenvalues of $U$. We arrange, by multiplying by an unimportant
overall phase, for this gap to be centred on the origin.

The idea now is to exploit the existence of the gap to provide a
more useful series representation for $H$. We do this by calculating
the fourier coefficients $d_k$ of the sawtooth wave $f(\theta)$
convolved with a sufficiently smooth smearing function
$\chi_\gamma(\theta)$; the fourier series then inherits a better
convergence from the smoothness properties of the smearing function.
That is, we define $d_k$ to be the fourier coefficients of
\begin{equation}
g(\theta) = (f\star \chi_\gamma)(\theta) = \int_{-\infty}^{\infty}
f(\theta-y)\chi_\gamma(y)dy.
\end{equation}
We choose $\chi_\gamma(y)$ to be a symmetric $C^\infty$ bump
function with compact support in the interval $[-\gamma, \gamma]$
(see the Appendix for further details.) Note that, as a consequence
of the compact support of $\chi_\gamma(y)$, $g(\theta) = f(\theta)$,
$\forall\theta \in (\gamma, 2\pi -\gamma)$. An application of the
convolution theorem then tells us that the fourier coefficients
$d_k$ are given by
\begin{equation}
d_k = \widehat{\chi}_\gamma(k) c_k,
\end{equation}
where $\widehat{\chi}_\gamma(\omega)$ is the \emph{fourier
transform} of $\chi_\gamma(y)$. (See Fig.~\ref{fig:smeartooth} for
an illustration of the smearing of the sawtooth wave.)

Using the fourier coefficients $d_k$ it is possible to construct a
logarithm $J$ of $U$ which is manifestly sparse if $U$ is. We begin
by constructing the following approximate hamiltonian:
\begin{equation}\label{eq:jfseries}
J = \sum_{k=-\infty}^{\infty}  \widehat{\chi}_\gamma(k) c_k U^k.
\end{equation}
Choosing $\gamma < \Delta$ allows us to conclude that, in fact, $H =
J$, because both $f(\theta)$ and $g(\theta)$ agree on the spectrum
of $U$.

Our final approximation $J_k$ to $H$ is defined by
\begin{equation}
J_k = \sum_{j=-k}^{k}  \widehat{\chi}_\gamma(j) c_j U^j.
\end{equation}
If $U$ is sparse, with only, say, polynomially many entries in $n$
in each row, then so is $U^j$ for $j$ constant. Thus, if we choose
$k$ to be a constant, then $J_k$ will only be polynomially less
sparse than $U$.

How big do we have to choose $k$? To see this we bound the
difference between $H$ and $J_k$ via an application of the triangle
inequality:
\begin{equation}
\|H-J_k\|_\infty \le \sum_{|j|>k} |\widehat{\chi}_\gamma(j)| |c_j|.
\end{equation}
Now, according to the properties of compactly supported $C^\infty$
bump functions described in the Appendix, $\widehat{\chi}_\gamma(j)$
has a characteristic width of $1/\gamma$, after which it decays
faster than any polynomial. Thus, choosing $k \gtrsim 1/(\epsilon^j
\Delta)$, for any $j\ge 1$, is sufficient to ensure that
$\|H-J_k\|_\infty$ can be made smaller than any prespecified
accuracy $\epsilon$.

Now to conclude, we define $V = e^{iJ_k}$ and use the upper bound
for $\|H-J_k\|_\infty$ which we've derived above to bound
$\|U-V\|_\infty$:
\begin{equation}
\|U-V\|_\infty \le \|H-J_k\|_\infty.
\end{equation}

\end{proof}

\begin{remark}
By choosing the smearing function $\chi_\gamma(y)$ to be a gaussian
a slightly better error scaling can be achieved at the expense of a
slightly more complicated argument: in this case $J$ doesn't equal
$H$ and one must bound the difference between them.
\end{remark}

\begin{example}
Consider the coined quantum walk on the ring of $n$ vertices: this
is the unitary matrix $U$ defined by $U = (|0\rangle\langle
0|\otimes \mathcal{T} + |1\rangle\langle 1|\otimes
\mathcal{T}^{\dag})H\otimes \mathbb{I}$, where $\mathcal{T}$ is the
unit translation operator $\mathcal{T}|j\rangle = |j+1\mod n\rangle$
and $H$ is the hadamard gate
$\frac{1}{\sqrt{2}}\left(\begin{smallmatrix} 1 & 1 \\ 1 &
-1\end{smallmatrix}\right)$. The spectrum of $U$ straightforward to
calculate using a fourier series \cite{ambainis:2001a}; one finds
that the eigenvalues $\lambda_k^\pm$ of $U$ are given by
\begin{equation}
\lambda_k^\pm = \frac{1}{\sqrt{2}}\cos\left(\frac{2\pi k}{n}\right)
\pm \frac{i}{\sqrt{2}}\sqrt{1+\sin^2\left(\frac{2\pi k}{n}\right)}.
\end{equation}
\end{example}
Clearly there is a gap $\Delta$ in the spectrum for all $n$
subtending an angle of $\theta$ with
\begin{equation}
\theta > 2\tan^{-1}(1) = \pi/2.
\end{equation}
Thus we find that there exists a logarithm $H$ of $U$ which can be
expressed as a sum of a few powers of $U$. Because $U$ is sparse, so
is $H$. (See Fig.~\ref{fig:hamiltonianme} for an illustration of the
logarithm of the coined quantum walk.)

\begin{remark}
The quantum fourier transform \cite{nielsen:2000a, preskillnotes} is
the unitary matrix $Q$ defined by the discrete fourier transform:
\begin{equation}
Q_{j,k} = \frac{1}{\sqrt{n}}e^{\frac{2\pi i jk}{n}}.
\end{equation}
The eigenvalues of $Q$ are well known: because $Q^4 = \mathbb{I}$
the eigenvalues are the fourth roots of unity. Thus $Q$ possesses a
gap of size $\Delta = \pi/2$ in its spectrum so we can construct a
logarithm $F$ of $Q$ as a series Eq.~(\ref{eq:jfseries}) in $Q$.
Although $F$ will be dense, it admits a description which is compact
(i.e., we can efficiently evaluate the matrix elements of $F$).
Given the logarithm $F$ it is straightforward to compute the square
root of $Q$: $\sqrt{Q} = e^{\frac{i}{2}F} = \sum_{j=0}^{\infty}
\frac{i^j}{2^jj!}F^j$.
\end{remark}

\begin{remark}
Our proof of Proposition~\ref{prop:logu} also holds for unitary
operators $U$ which are only \emph{approximately} $Z$-local, i.e.,
when the condition that $U_{j,k} = 0$ when $(j, k) \not\in E$ is
replaced with $U_{j,k} \le e^{-\kappa \dist(j,k)}$, or similar.
\end{remark}

The are several questions left open at this point. Perhaps most
interesting is the question of how to provide a combinatorial
characterisation of unitary operators which possess a gap in their
spectrum. Presumably such a characterisation would take the form of
a necessary condition, not unlike the isoperimetric inequality
\cite{chung:1997a}.

\begin{acknowledgments}
Many thanks to Jens Eisert for providing me with numerous helpful
comments, and for many enlightening and inspiring conversations.
Thanks also, of course, to Scott Aaronson for helpful
correspondence, discussions, and for suggesting this problem in the
first place! This work was supported, in part, by the Nuffield
foundation.
\end{acknowledgments}

\appendix

\section{Properties of smooth cutoff functions}\label{app:cutoff}


In this Appendix we briefly review the properties of compactly
supported $C^\infty$ cutoff functions.

Of fundamental utility in our derivations is a class of functions
known as \emph{compactly supported $C^\infty$ bump functions}. These
functions are defined so that their fourier transform
$\widetilde{\chi}_\gamma(\omega)$ is compactly supported on the
interval $[-\gamma, \gamma]$, and equal to $1$ on the middle third
of the interval. Such functions satisfy the following derivative
bounds
\begin{equation}\label{eq:chiderbound}
\frac{d^j\widehat{\chi}_\gamma(\omega)}{d\omega^j} \lesssim
\gamma^{-j},
\end{equation}
for all $j$ with the implicit constant depending on $j$. (If we have
two quantities $A$ and $B$ then we use the notation $A\lesssim B$ to
denote the estimate $A\le CB$ for some constant $C$ which only
depends on unimportant quantities.) This is just about the best
estimate possible given Taylor's theorem with remainder and the
constraints that $\widehat{\chi}_\gamma(\omega)$ is equal to $1$ at
$\omega = 0$ and $\widehat{\chi}_\gamma(\omega)$ is compactly
supported.

The function $\chi_\gamma(t)$ has support throughout $\mathbb{R}$
but it is decaying rapidly. To see this consider
\begin{equation}
\chi_\gamma(t) = -\frac{1}{2\pi}\int_{-\infty}^{\infty} \frac{1}{it}
e^{-it\omega}\frac{d}{d\omega}\widehat{\chi}_\gamma(\omega) d\omega
\end{equation}
which comes from integrating by parts. Continuing is this fashion
allows us to arrive at
\begin{equation}
\chi_\gamma(t) = \frac{1}{2\pi}\int_{-\infty}^{\infty}
\left(-\frac{1}{it}\right)^j
e^{-it\omega}\frac{d^j}{d\omega^j}\widehat{\chi}_\gamma(\omega)
d\omega
\end{equation}
Since $\widehat{\chi}_\gamma(\omega)$ has all its derivatives
bounded, according to (\ref{eq:chiderbound}), and using the compact
support of $\widehat{\chi}_\gamma(\omega)$ we find
\begin{equation}
\begin{split}
|\chi_\gamma(t)| &\lesssim \left|\int_{-\gamma}^{\gamma}
\left(\frac{1}{it}\right)^j e^{-it\omega} \gamma^{-j} d\omega\right|
\\
&\lesssim \int_{0}^{\gamma} \frac{1}{|\gamma t|^j} d\omega \\
&\lesssim \frac{1}{\gamma^{j-1}|t|^j},
\end{split}
\end{equation}
for all $j\in \mathbb{N}$. Thus we find that $\chi_{\gamma}(t)$
decays to $0$ faster than the inverse of any polynomial in $t$ with
characteristic ``width'' $1/\gamma$. The existence and construction
of such functions is discussed, for example, in \cite{vaaler:1981a,
vaaler:1985a}.


\begin{thebibliography}{17}
\bibitem{endnote19}{It is entirely reasonable that several
definitions of nonlocality should be allowed: if we demand some
geometric constraints on the particles the gates are allowed to act
on, such as only on nearest-neighbouring spins, then the nonlocality
cost would be different.}
\bibitem{endnote20}{$\delimiter "426830A j|U^l|k \delimiter "526930B $
can only be nonzero if there exists a path of length $\le l$ between
vertices $j$ and $k$.}
\end{thebibliography}
\end{document}